\documentclass[prc,preprint,showpacs,showkeys,tightenlines,%
nofootinbib]{revtex4}

\usepackage{graphicx}  
\usepackage{bm}  

\newcommand{\beq}{\begin{equation}}
\newcommand{\eeq}{\end{equation}}
\newcommand{\bea}{\begin{eqnarray}}
\newcommand{\eea}{\end{eqnarray}}

\newcommand{\kf}{k_F}

\newcommand{\kt}{\widetilde k}

\def\vec#1{{\bf #1}}

\newcommand{\nab}{\overrightarrow{\nabla}}
\newcommand{\nabsq}{\overrightarrow{\nabla}^{2}\!}

\newcommand{\galnab}{\tensor{\nabla}}

\begin{document}

\title{Quasiparticle Properties in Effective Field Theory}
\author{L. Platter}\email{l.platter@fz-juelich.de}
\affiliation{Forschungszentrum J{\"u}lich, Institut f{\"ur} Kernphysik,
    D-52425 J{\"u}lich, Germany}

\author{H.-W. Hammer}\thanks{
Address after October 1, 2002: Institut f\"ur Theoretische Physik, 
  Karl-Franzens-Universt\"at Graz, A-8010 Graz, Austria}
\affiliation{Department of Physics,
         The Ohio State University, Columbus, OH\ 43210, USA}

\author{Ulf-G. Mei{\ss}ner}\email{u.meissner@fz-juelich.de}
\affiliation{Forschungszentrum J{\"u}lich, Institut f{\"ur} Kernphysik,
    D-52425 J{\"u}lich, Germany\\
Institut f\"ur Theoretische Physik, 
  Karl-Franzens-Universt\"at Graz, A-8010 Graz, Austria 
  \vskip 0cm}

%
\date{August, 2002}

\begin{abstract}
The quasiparticle concept is an important tool for the
description of many-body systems. We study the
quasiparticle properties for dilute Fermi systems
with short-ranged, repulsive interactions using effective field theory.
We calculate the proper self-energy contributions at order $(\kf/\Lambda)^3$,
where $\Lambda$ is the short-distance scale that sets the size
of the effective range parameters and $\kf$ the Fermi momentum.
The quasiparticle energy, width, and effective mass
to ${\cal O}((\kf/\Lambda)^3 )$ are derived from the calculated self-energy.
\end{abstract}

\smallskip
\maketitle

\section{Introduction}

A relatively new approach to many-body phenomena is that of effective
field theory (EFT). The EFT approach is designed to
exploit the separation of scales in physical
systems \cite{WEINBERG79,LEPAGE89,KAPLAN95}.
Only low-energy (or long-range) degrees of freedom are included
explicitly, with the rest parametrized in terms of the most general
local (contact) interactions.  Using renormalization,
the influence of high-energy states on low-energy observables
is captured in a small number of low-energy constants.
Thus, the EFT describes universal low-energy physics independent of
detailed assumptions about the short-distance dynamics. Furthermore,
the application of EFT methods to many-body problems gives
a consistent organization of many-body corrections, with reliable error
estimates, and insight into the analytic structure of observables.
EFT provides a model independent description of finite density
observables in terms of parameters that can be fixed from
scattering in the vacuum. In low-energy particle and nuclear 
physics, EFT methods have successfully been applied
to mesonic systems \cite{GaL85}, processes with one baryon
\cite{Mei00}, and more recently also to processes involving two
and more baryons \cite{BEANE99,Bedaque:2002mn}.

A description in terms of quasiparticles is central to many approaches to
the many-body problem. It has proven useful
in many physical systems from condensed matter physics to nuclear 
physics \cite{NEGELE88,FETTER71}.
If the states of the noninteracting system are gradually transformed into 
the interacting states when the 
interactions are switched on adiabatically, the excitations of the
interacting system near the Fermi surface (the quasiparticles)
have the same character as the excitations of the noninteracting 
Fermi gas. This is the basic assumption underlying Landau's theory of 
Fermi liquids \cite{Landau56,Noz97}. The assumption holds, e.g.,
for a dilute Fermi system with repulsive, short-ranged interactions.
This system has been a benchmark problem for many-body techniques for 
a long time \cite{efimov,baker,bishop,HAMMER00}. 
In particular, the low density allows for a controlled 
expansion in $\kf a_s$ where $\kf$ is the Fermi momentum and $a_s$ is 
the S-wave scattering length of the fermions.
In the framework of EFT, this corresponds to an expansion
in $\kf/\Lambda$ where $\Lambda$ is the breakdown scale of the EFT.
For a natural system the scale $\Lambda$ sets the scale of $a_s$
and all other effective range parameters. Initially, 
there were not many applications for this model system, but recently 
ultra-cold Fermi gases of atoms have been achieved \cite{jin99,hulet01}.
The dilute Fermi gas also serves as a simpler test case on the way
to an EFT description of nuclear matter. 
The quasiparticle properties in the dilute Fermi gas to
${\cal O}((\kf/\Lambda)^2)$ were first calculated by Galitskii
\cite{GALITSKII}. An incomplete numerical study  of the quasiparticle 
properties to ${\cal O}((\kf /\Lambda)^3)$ was  
carried out in Ref.~\cite{bund}.
In Ref.~\cite{SARTOR80}, the proper self-energy was calculated off-shell
to ${\cal O}((\kf/\Lambda)^2)$ and the momentum dependence of the
effective mass was studied in detail.

In this paper, we perform the first complete calculation of
the quasiparticle properties in the dilute Fermi
gas to ${\cal O}((\kf/\Lambda)^3)$ in the low-density expansion. 
At this order both the S-wave effective range $r_s$ and the P-wave scattering 
length $a_p$ contribute. We calculate
the quasiparticle energies $\epsilon_k$, width $\gamma_k$, and 
effective mass $M^*$. As a check, we also calculate the chemical potential
$\mu$ and the energy per particle and compare with the known 
results \cite{efimov,baker,bishop,HAMMER00}. 
The outline of this paper is as follows. In Sec.~\ref{sec_EFT} we 
repeat briefly the elements of an EFT approach to dilute Fermi systems. 
In Sec.~\ref{sec_selfenergy}, we calculate the proper self-energy 
at order $(k_F/\Lambda)^3$. The quasiparticle properties to 
${\cal O}((k_F/\Lambda)^3)$ are derived from the self-energy 
in Sec.~\ref{sec_quasip}. In Sec.~\ref{sec_summary}, we summarize and
present our conclusions.

\section{Effective Field Theory for Dilute Fermi Systems}
\label{sec_EFT}
In this section, we review briefly the EFT approach for dilute Fermi 
systems. We will restrict ourselves to repulsive interactions. 
An EFT for such a system is discussed in detail in Ref.~\cite{HAMMER00}.
In contrast to more traditional many-body approaches 
\cite{efimov,baker,bishop} the EFT for the dilute Fermi gas
is fully perturbative.  
It is appropriate to consider a natural EFT for heavy, nonrelativistic 
fermions of mass M with spin-independent interactions whose strength 
is correctly determined by naive dimensional analysis.
The scale of all effective range parameters is set by $1/\Lambda$, 
where $\Lambda$ is the breakdown scale of the theory. 
For momenta $ k \ll \Lambda$, all interactions appear short-ranged
and can be modeled by contact terms. Therefore,
we consider a local Lagrangian for a nonrelativistic fermion
field that is invariant under Galilean, parity, and time-reversal
transformations:
\bea
  {\cal L}  &=&
       \psi^\dagger \biggl[i\partial_t + \frac{\nabsq}{2M}\biggr]
                 \psi - \frac{C_0}{2}(\psi^\dagger \psi)^2
            + \frac{C_2}{16}\Bigl[ (\psi\psi)^\dagger
                                  (\psi\galnab^2\psi)+\mbox{ H.c.}
                             \Bigr]  
\nonumber\\  &&
         + \frac{C_2'}{8} (\psi \galnab \psi)^\dagger \cdot
             (\psi\galnab \psi)    
         +  \ldots
\,,\label{lag}
\eea
where $\galnab=\overleftarrow{\nabla}-\nab$ is the Galilean invariant  
derivative and H.c.\ denotes the Hermitian conjugate.
Higher-order time derivatives are not included, since it is most
convenient for our purposes to eliminate them in favor of spatial gradients.

For convenience,
we choose dimensional regularization with minimal subtraction as 
our renormalization scheme. Calculating the two-particle scattering
amplitude in the vacuum to ${\cal O}((k/\Lambda)^3)$, 
the coefficients $C_{2i}$ can be 
determined by matching to the leading coefficients of the effective range 
expansion (see, e.g.,  Ref.~\cite{HAMMER00} for details). Thus, 
the $C_{2i}$ can be expressed completely in terms of the
effective range parameters:
\begin{equation}
  C_0 =\frac{4 \pi a_s}{M},\qquad C_2=C_0 \frac{a_s r_s}{2}, \qquad 
\mbox{and}\qquad C_2^{\prime}=\frac{4 \pi a_p^3}{M},
\end{equation} 
where $a_s$, $a_p$ and $r_s$ are the S- and P-wave scattering lengths 
and S-wave effective range, respectively.

We now turn to the finite density system. 
First, consider a noninteracting Fermi gas. The single particle states of 
the system are characterized by a momentum $k$ and a spin quantum
number $s$. In particular, the ground-state of the system is the state 
in which all single-particle states with momentum less than the Fermi
momentum $k_F$ are occupied and all other single-particle states are 
empty. Any excited state of the system can be created by removing particles 
with momentum less than $k_F$ and adding particles with momentum greater 
than $k_F$.

Now we turn on the interactions. We assume that there is a one-to-one 
correspondence between the states of the noninteracting Fermi gas and 
those of the interacting system if the interaction is switched on 
adiabatically. Note that this is not the case if the 
interaction is attractive. In this case, it is energetically favorable 
to form Cooper pairs and our calculation does not lead to the correct
ground state. 
The elementary excitations of the interacting system 
correspond thus to the particle and hole excitations of the 
noninteracting Fermi gas,
and are referred to as quasiparticles and quasiholes, respectively. 
Information about the properties of quasiparticles
is contained in the proper self-energy $\Sigma^{*}(k_0,\vec{k})$. 
In terms of diagrams, the proper self-energy represents 
one-particle irreducible and amputated self-energy insertions. 
The proper self-energy is related to the full Green's 
function $G_{\alpha\beta}(\tilde k)$ by Dyson's equation \cite{FETTER71}:
\begin{equation}
G_{\alpha \beta}(\tilde k)=G^{0}_{\alpha \beta}(\tilde k)+
G^{0}_{\alpha\lambda}(\tilde k)\Sigma^{*}(\tilde k)_{\lambda\mu}
G_{\mu\beta}(\tilde k)\,,
\end{equation}
where ${\tilde k}\equiv (k_0,\vec{k})$, $\alpha,\beta,...$ are 
spin indices, and the subscript $0$ denotes 
the noninteracting
Green's function defined in Eq.~(\ref{freeprop_fd}) below.
For spin-independent interactions this equation can be solved 
analytically to give:
\beq
G_{\alpha\beta}(\tilde k)=\frac{\delta_{\alpha\beta}}
{k_0-\vec{k}^2/(2M) -\Sigma^{*}(\tilde k)}\,.
\label{greens_full}
\eeq

\begin{figure}[b]
\centerline{\includegraphics*[width=4.5in,angle=0]{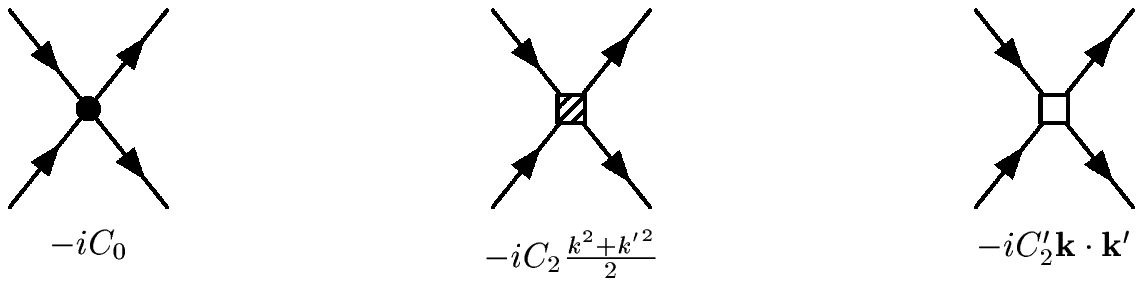}} 
\caption{\label{fig_vertex}Feynman rules for the vertices arising 
 from the Lagrangian (\ref{lag}).
 The relative momenta of the incoming and outgoing particles,
 are $2\vec{k}$ and $2\vec{k}'$, respectively. 
 The spin indices have been suppressed.}
\end{figure}

Here, we only repeat the Feynman rules for the proper self-energy:
The vertices that follow from the 
Lagrangian (\ref{lag}) are illustrated in Fig.~\ref{fig_vertex}.
For short-range interactions the direct and exchange
contributions differ only by a spin degeneracy factor. As a consequence,
it is convenient to use so-called \lq\lq Hugenholtz'' diagrams where
the direct and exchange contributions are combined in local four-fermion
vertices.
The Feynman rules are \cite{HAMMER00}:
\begin{enumerate}
\item 
Write down all Hugenholtz diagrams that scale with a given order 
in $\kf$ as determined by the power counting rules given below.
\item 
Assign nonrelativistic four-momenta (frequency and three-momentum)
to all lines and enforce four-momentum conservation
at each vertex.
\item
For each vertex, include the corresponding expression from
Fig.~\ref{fig_vertex}.
For spin-independent interactions, the two-body vertices have
the structure $(\delta_{\lambda\gamma}\delta_{\beta\delta}
+\delta_{\lambda\delta}\delta_{\beta\gamma})$,
where $\lambda,\beta$ are the spin indices of the incoming lines     
and $\gamma,\delta$ are the spin indices of the outgoing 
lines.\footnote{It is most convenient 
to take care of the antisymmetrization in the spin summations. This is
discussed in rule \ref{spinsum}.} For each internal line include a factor
$iG_0 (\kt)_{\lambda\gamma}$, where
$\kt \equiv (k_0,\vec{k})$ is the four-momentum assigned to the line,
$\lambda$ and $\gamma$ are spin indices, and
\beq
    iG_0 (\kt)_{\lambda\gamma}=i\delta_{\lambda\gamma}
    \left( \frac{\theta(k-\kf)}{k_0-\vec{k}^2/(2M)+i\epsilon}
      +\frac{\theta(\kf-k)}{k_0-\vec{k}^2/(2M)-i\epsilon}\right) \ .
      \label{freeprop_fd}
\eeq
\item Perform the spin summations in the diagram.
In every closed fermion loop, substitute a spin degeneracy
factor $-g$ for each $\delta_{\lambda\lambda}$.
\label{spinsum}
\item Integrate over all independent momenta with a factor       
$\int\! d^4 k /(2\pi)^4$
where $d^4k\equiv dk_0\,d^3k$.
If the spatial integrals are divergent, they are defined in $D$ spatial
dimensions and renormalized using minimal subtraction as discussed in
Ref.~\cite{HAMMER00}. For a tadpole line with four-momentum $\tilde{k}$,
multiply by $\exp(ik_0\eta)$ and take the limit $\eta\to 0^+$
after the contour integrals have been carried out. This procedure
automatically takes into account that such lines must be hole lines.
\item Multiply every Hugenholtz diagram by the appropriate 
 symmetry factor.\footnote{This symmetry factor is most easily
obtained by resolving the Hugenholtz vertices in Fig.~\ref{fig_vertex}
into vertices with direct and exchange contributions. For a Hugenholtz
diagram with $n$ vertices this substitution generates $2^n$ new
diagrams. Now identify the symmetry factor that corrects for the double 
counting of topologically equivalent diagrams among the $2^n$ diagrams.}
\end{enumerate}
Furthermore, we repeat the power-counting rules that determine the 
power of $\kf$ with which a given diagram scales:
\begin{itemize}
\item for every propagator a factor $M/k_F^2$~,
\item for every loop integration a factor of $k_F^5/M$~,
\item for every n-body vertex with $2i$ derivatives a factor 
$k_F^{2i}/(M\Lambda^{2i+3n-5})$~,
\end{itemize}
where $\Lambda$ is the breakdown scale of the theory which is 
determined by the physics not included in the
EFT (such as the mass of a heavy exchanged particle). 
Thus, $k_F$ has to be much smaller than $\Lambda$, which is 
just the case for low density systems with short-range interactions 
where $\rho\propto k_F^3\ll 1/a_s^3\sim\Lambda^3$. In the above power
counting rules, the external momentum $\vec{k}$
in $\Sigma^*$ is also counted as order $\kf$.
This is appropriate, since the external momentum is converted to 
$\kf$ if observables like the energy density or quasiparticle properties
are calculated. All proper self-energy diagrams have an overall factor 
of the Fermi energy $\kf^2/(2M)$
which we suppress in the labeling of different orders. The leading 
order contribution then starts at ${\cal O}(\kf/\Lambda)$. As a
consequence, the expansion of $\Sigma^*$ reads:
\beq
\Sigma^*(k_0,\vec{k})=\Sigma^*_1 + \Sigma^*_2 (k_0,\vec{k})
+ \Sigma^*_3 (k_0,\vec{k}) + \ldots\,,
\label{Sigmaexp}
\eeq
where the lower index indicates the corresponding power of $\kf/\Lambda$.
In a homogeneous system $\Sigma^*(k_0,\vec{k})$ depends only on 
$k=|\vec{k}|$.
The expressions for the self-energy through ${\cal O}((\kf/\Lambda)^2)$ 
for $k_0=k^2/(2M)$ (sometimes called \lq\lq on shell'')
were first calculated by Galitskii \cite{GALITSKII}.
The general expressions for arbitrary $k_0$ were obtained
in Ref.~\cite{SARTOR80}. For completeness, we quote the
on-shell expressions for $\Sigma^*_1$ and $\Sigma^*_2$ which
will be needed later on:
\begin{eqnarray}
&&\Sigma^*_1 =
\frac{\kf^2}{2M} (g-1) \frac{4}{3\pi} (\kf a_s)\,,  \\
&&{\rm Re}\,\Sigma^*_2(\mbox{$\frac{k^2}{2M}$},k) =\frac{k_F^2}{2M}
(g-1)(k_F a_s)^2\frac{4}{15\pi^2}\frac{1}{v}\biggl[11v +
2 v^5\ln \biggl|\frac{v^2}{v-1}\biggr|\nonumber\\
&&\quad\qquad-10(v^2-1)\ln\biggl|\frac{v+1}{v-1}\biggr|-(2-v^2)^{5/2}
\ln\biggl|\frac{1+v\sqrt{2-v^2}}{1-v\sqrt{2-v^2}}\biggr|\biggr]\,,
\\
0<k<k_F:\,&&{\rm Im}\,
\Sigma^*_2(\mbox{$\frac{k^2}{2M}$},k) =
(g-1)\frac{k_F^2}{4 \pi M}(k_F a_s )^2 (1-v^2)^2
\label{eq:Imsig21}\,,\\
k_F<k<\sqrt{2} k_F:\,&&{\rm Im}\Sigma^*_2(\mbox{$\frac{k^2}{2M}$},k) 
=(1-g)\frac{k_F^2}{15\pi M}(k_F a_s)^2 \frac{1}{v}
\biggl(5v^2-7+2(2-v^2)^{5/2}\biggr)\,,
\label{eq:Imsig22}
\end{eqnarray}
with $v=k/k_F$.
If the imaginary parts in Eqs.~(\ref{eq:Imsig21}) and (\ref{eq:Imsig22})
are expanded around the Fermi surface at $v=1$, the first nonvanishing
contribution occurs at ${\cal O}((1-v)^2)$ as required by Luttinger's 
general theorem \cite{LUTTINGER61}.
In the next section, we calculate the proper self-energy 
contribution $\Sigma^*_3$ at ${\cal O}((\kf/\Lambda)^3)$ for 
$k_0=k^2/(2M)$. As we will show in Section \ref{sec_quasip},
only the on-shell value of $\Sigma^*_3$ is required
for the quasiparticle properties.

\section{Proper Self-Energy at ${\cal O}((\kf/\Lambda)^3)$}
\label{sec_selfenergy}
In this section, we calculate the contribution to the proper
self-energy at ${\cal O}((\kf/\Lambda)^3)$: $\Sigma^*_3$. 
\begin{figure}[t]
\centerline{\includegraphics*[width=4in,angle=0]{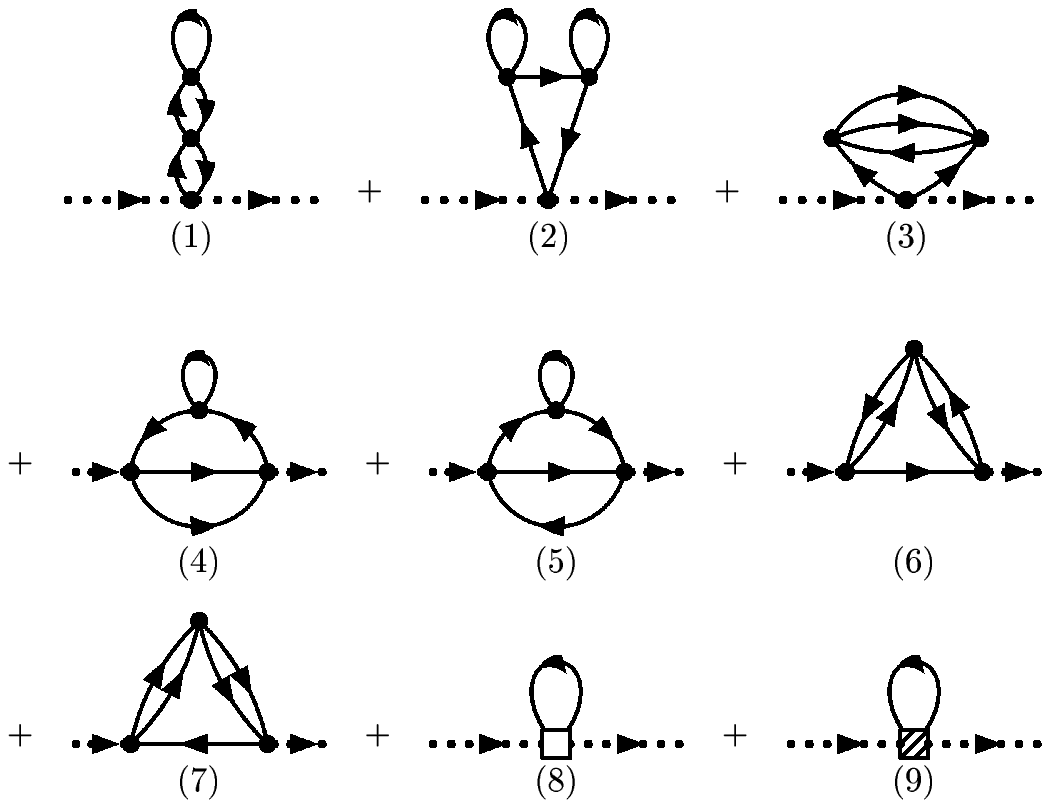}} 
\caption{\label{fig_sigma3}
Hugenholtz diagrams contributing to $\Sigma^{*}_{3}(k_0,k)$.
Solid lines indicate noninteracting fermion propagator (\ref{freeprop_fd}),
while dotted lines indicate amputated fermion propagators.
The expressions for the vertices are given in Fig.~\ref{fig_vertex}.
}
\end{figure}  
$\Sigma^{*}_{(3)}(k_0,k)$ is given by the nine diagrams in shown 
Fig.~\ref{fig_sigma3}. We will omit the calculational details 
in this section and show a sample calculation of diagram (7)
in Appendix \ref{app_integration}.
Diagrams (1) to (3) vanish when evaluated. Diagrams (4) and (5) 
can be related to the derivative of $\Sigma^*_2 (k_0,k)$ with
respect to $k_0$:
\beq
\Sigma^*_{3(4)}(k_0,k)+ \Sigma^*_{3(5)} (k_0,k)=
-\Sigma^*_1 \left. \frac{\partial \Sigma^*_2 (k_0,k)}{\partial k_0}
\right|_{k_0=\frac{k^2}{2M}} \,.
\eeq
They do not contribute the quasiparticle properties at 
${\cal O}((\kf/\Lambda)^3)$ as will be shown in the next section.
Diagrams (8) and (9) can be evaluated analytically and give purely 
real results. For  diagrams (6) and (7), 
we could not obtain analytical results; instead we have
simplified the analytical expressions as far as possible and
calculated them numerically using the Vegas Monte Carlo integration
routine \cite{vegas}. 
Below, we give parametrizations for the contributions of diagrams (6) and 
(7) to $\Sigma^*_3$ from a fit to the numerical results. We have fitted
the results for different external momenta $k$ to 
polynomials in $(\kf-k)$ in an expansion around the Fermi surface.
We have truncated the expansion at the order in $(\kf-k)$ at which
adding another term does not improve the $\chi^2$ of the fit anymore.
According to this criterion a polynomial of seventh degree is sufficient 
for the real parts, while the imaginary parts require a polynomial 
of ninth degree in $(\kf-k)$.  
The results for the contributions to the proper self-energy 
at ${\cal O}((\kf/\Lambda)^3)$ are:
\begin{eqnarray}
\nonumber \hbox{Re}\,\Sigma^{*}_{3(6)}(\mbox{$\frac{k^2}{2M}$},k)
&=&\frac{k_F^2}{2 M}(g-1)(g-3) \frac{2^3}{\pi^3}(k_F a_s)^3~
    \biggl\{0.5908-0.8829(1-v)\\
&&\qquad +0.2140(1-v)^2 +3.7284(1- v)^3-2.7392(1- v)^4\nonumber\\
&&\qquad -4.7509(1-v)^5+7.0977(1-v)^6-2.6709(1-v)^7\biggr\}\,,
\label{eq:para1}
\\[0.3cm]
\nonumber \hbox{Re}\,\Sigma^{*}_{3(7)}(\mbox{$\frac{k^2}{2M}$},k)
&=&\frac{k_F^2}{2 M}(g-1)\frac{2^3}{\pi^3}(k_F a_s)^3
\biggl\{0.7811 +1.0589(1-v)\nonumber\\
&&\qquad+0.4160(1-v)^2+1.9942(1-v)^3-3.1936(1-v)^4\nonumber \\
&&\qquad-4.8352(1-v)^5+9.6964(1-v)^6 -4.1846(1-v)^7
\biggr\}~,
\label{eq:para2}\\[0.3cm]
\hbox{Re}\,\Sigma^{*}_{3(8)}(\mbox{$\frac{k^2}{2M}$},k)&=&\frac{k_F^2}{2 M}
\frac{(g+1)}{\pi}(k_F a_p)^3\biggl\{\frac{1}{5}+\frac{1}{3}v^2
\biggr\}\,,\\[0.3cm]
\hbox{Re}\,\Sigma^{*}_{3(9)}(\mbox{$\frac{k^2}{2M}$},k)
&=&\frac{k_F^2}{2 M}\frac{(g-1)}{2\pi}\,k_F^3 a_s^2 r_s\,
\biggl\{\frac{1}{5} +\frac{1}{3}v^2\biggr\}\,,\\[0.3cm]\nonumber
\hbox{Im}\,\Sigma^{*}_{3(6)}(\mbox{$\frac{k^2}{2M}$},k)
&=&\frac{k_F^2}{2 M }\frac{16(g-1)(g-3)}{\pi^2}(k_F a_s)^3
\biggl[\biggl\{-0.0065(1- v)^2-1.0778(1- v)^3\nonumber\\
&&\;+5.2913(1-v)^4-13.0528(1-v)^5+18.7234(1-v)^6\nonumber \\
&&\;-15.4468(1-v)^7+6.7036(1-v)^8-1.1653(1-v)^9
\biggr\}\theta(1-v)\nonumber \\
&&\;+\biggl\{-0.0700(1-v)^2 -5.3866(1-v)^3-84.0177(1-v)^4
\nonumber \\
&&\;-721.4519(1- v)^5-3546.7045(1-v)^6-9902.6272(1-v)^7
\nonumber \\
&&\;-14617.8867(1-v)^8-8855.7811(1-v)^9\biggr\}
\theta(v-1)\biggr]\,,
\label{eq:para3}
\\[0.3cm]\nonumber
\hbox{Im}\,\Sigma^{*}_{3(7)}(
\mbox{$\frac{k^2}{2M}$},k)&=&\frac{k_F^2}{2 M}
\frac{8(g-1)}{\pi^2}(k_F a_s)^3
\biggl[\biggl\{-0.3835(1-v)^2+3.1281(1-v)^3\nonumber \\
&&\;-7.1868(1-v)^4 +12.1392(1-v)^5-16.3114(1-v)^6\nonumber \\
&&\;+14.4972(1-v)^7-7.2813(1-v)^8 +1.5639(1-v)^9
\biggr\}\theta(1-v)\nonumber \\
&&\;+ \biggl\{0.3715(1-v)^2+ 8.0530(1- v)^3 +77.9691(1-v)^4
\nonumber \\
&&\;+532.4357(1-v)^5+2237.8794(1-v)^6+5462.8759(1-v)^7
\nonumber \\
&&\;+7100.2592(1-v)^8+3791.8271(1-v)^9
\biggr\}\theta(v-1)\biggr]~,
\label{eq:para4}
\\[0.3cm] \nonumber
\hbox{Im}\,\Sigma^{*}_{3(8)}(\mbox{$\frac{k^2}{2M}$},k)&=&0\,,\\[0.3cm]
\hbox{Im}\,\Sigma^{*}_{3(9)}(\mbox{$\frac{k^2}{2M}$},k)&=&0\,,
\end{eqnarray}
where $v=k/\kf$. In Eqs.~(\ref{eq:para1}), (\ref{eq:para2}), (\ref{eq:para3}),
and (\ref{eq:para4}), we give the numerical constants to four decimal places
to ensure that the parametrizations differ from the exact numerical values 
by less than $1\%$.
Also, note that the parametrizations for the imaginary parts start with
the $(1-v)^2$ term and satisfy Luttinger's theorem \cite{LUTTINGER61} by 
construction.

\section{Quasiparticle Properties}
\label{sec_quasip}
The elementary excitations of the interacting system correspond to the 
particle and hole excitations of the noninteracting Fermi gas,
and are referred to as quasiparticles and quasiholes, respectively. 
Information about the properties of quasiparticles
is contained in the proper self-energy $\Sigma^{*}(k_0,k)$. 
The singularities of the full Green's function (\ref{greens_full})
determine the excitation energies $\epsilon_k$ of the system and their 
corresponding widths $\gamma_k$. In the quasiparticle approximation,
the full Green's function (\ref{greens_full}) can be written as
\beq
G(k_0,k) = \frac{Z}{k_0 -\epsilon_k-i\gamma_k} \quad +\quad
\mbox{regular terms}\,,
\eeq
where $Z$ is the wave function renormalization.
To determine $\epsilon_k$ and $\gamma_k$, we have to solve the equation:
\beq
\epsilon_k+i\gamma_k-\frac{k^2}{2M}-\Sigma^{*}(\epsilon_k+i\gamma_k,
  k)=0\,,
\label{quasipoles}
\eeq
while the wave function renormalization $Z$ is given by
\beq
\label{eq:wfrenorm}
Z^{-1}=1-\left.\frac{\partial \Sigma^*(k_0,k)}{\partial k_0}
\right|_{k_0=\epsilon_k+i\gamma_k,\,k=\kf} \,.
\eeq
The effective mass $M^{*}$ of the quasiparticle
is defined using the group velocity of a 
quasiparticle state at the Fermi surface; that is the slope of the 
excitation energy at $p=k_F$ \cite{FETTER71}:
\beq
M^{*}=k_F \biggl(\frac{\partial \epsilon_k}{\partial k}\biggl|_{k=k_F}
\biggr)^{-1}~.
\eeq
The effective mass can be extracted from the heat capacity of the 
system in the zero temperature limit \cite{FETTER71}:
\beq
\frac{C_V}{V}=\frac{k_B^2 T M^{*}k_F}{3} \quad \hbox{as} \quad 
T\rightarrow 0~.
\eeq
Various other observables can be computed from these results as well. 
The chemical potential $\mu$ describes the minimal 
energy required to add an additional particle
to a system. Consequently, it is given by the quasiparticle excitation 
energy at the Fermi surface:
\beq\label{chemical}
 \mu=\left.\epsilon_{k}\right|_{k=\kf}\,.
\eeq
Another observable, which allows a simple check of the calculated
excitation energies, is the energy per particle. One can compute the energy 
per particle directly from the proper self-energy $\Sigma^{*}(p_0,p)$. 
It is easier, however, to use thermodynamical identities. The chemical 
potential at zero entropy $S$ is related to the exact 
ground-state energy $E$ by the thermodynamic identity:
\beq
\mu=\biggl(\frac{\partial E}{\partial N}\biggr)_V \quad \hbox{at} \quad S=0~.
\eeq
Integrating this equation at a constant volume $V$ and zero 
entropy gives:
\beq
\label{ground-state}
E=\int_0^N\hbox{d}N^\prime \mu (S=0,V,N^\prime)~.
\eeq
Since the $N$ dependence of $\mu$ is hidden in the Fermi momentum $k_F$ 
through the relation $k_F=(6\pi^2N/gV)^{1/3}$,
the energy per particle is easily evaluated by multiplying each term  
in the chemical potential of order $(k_F)^\lambda$ by a factor of 
$3/(3+\lambda)$.

\begin{figure}[ht]
\centerline{\includegraphics*[width=4in,angle=0]{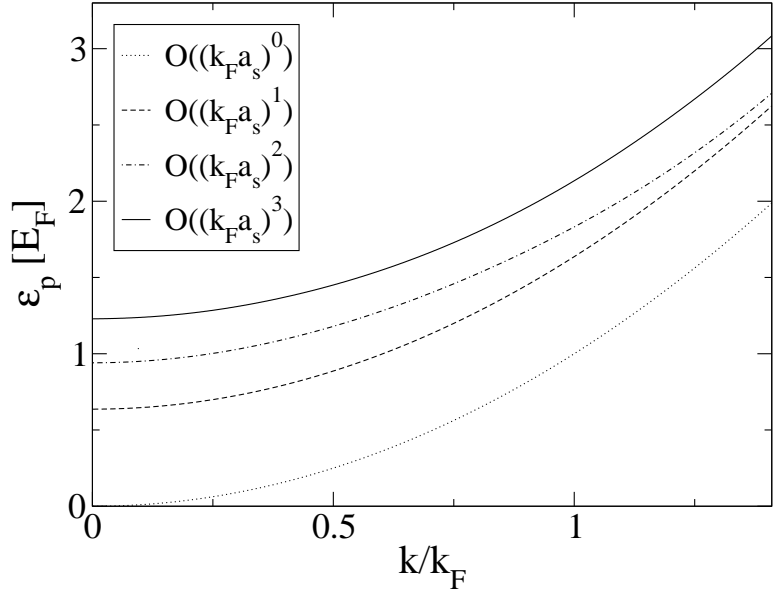}} 
\caption{\label{fig_epsk}
The quasiparticle energy $\epsilon_p$ in units of the Fermi energy
$E_F$ up to order $(k_F/\Lambda)^0$ (dotted line),
$(k_F/\Lambda)^1$ (dashed line), $(k_F/\Lambda)^2$ (dash-dotted line),
and  $(k_F/\Lambda)^3$ (solid line) for $\kf a_s =k_F a_p=k_F r_s= 0.5$ 
and $g=4$.}
\end{figure}

\begin{figure}[ht]
\centerline{\includegraphics*[width=4.2in,angle=0]{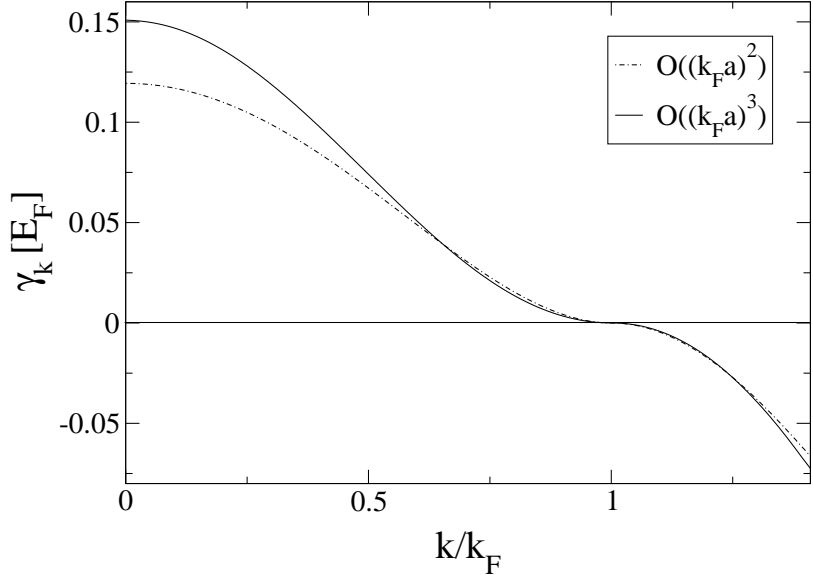}\ } 
\caption{\label{fig_gamk} 
The quasiparticle width $\gamma_p$ in units of the Fermi energy
$E_F$ up to order $(k_F/\Lambda)^2$ (dash-dotted line) and up to order 
$(k_F/\Lambda)^3$ (solid line) for $\kf a_s =k_F a_p=k_F r_s= 0.5$ and $g=4$.
}
\end{figure}

To obtain all these quantities, we solve Eq.~(\ref{quasipoles})
perturbatively in $\kf/\Lambda$.
We therefore expand $\epsilon_k$ and  $\gamma_k$ in powers of
$\kf/\Lambda$ as well:
\begin{eqnarray}
\epsilon_k &=& \epsilon_{k,0} + \epsilon_{k,1} + \epsilon_{k,2} + \ldots
\nonumber \,,\\
\gamma_k &=& \gamma_{k,0} + \gamma_{k,1} + \gamma_{k,2}  + \ldots \,.
\label{quasipropexp}
\end{eqnarray}
Inserting Eqs.~(\ref{Sigmaexp}) and (\ref{quasipropexp}) into
Eq.~(\ref{quasipoles}) and requiring Eq.~(\ref{quasipoles}) to
hold for every order in $\kf/\Lambda$, we obtain the matching 
equations. The solutions through order $(\kf/\Lambda)^2$ were first
obtained by Galitskii \cite{GALITSKII} (see also Ref.~\cite{FETTER71}).
For completeness, we quote their results:
\begin{eqnarray}
\epsilon_{k,0} = \frac{k^2}{2M}\,,&\qquad& \gamma_{k,0} = 0\,,
\nonumber \\
\epsilon_{k,1} = \Sigma^*_1
\,,&\qquad& \gamma_{k,1} = 0\,,
\nonumber \\
\epsilon_{k,2} = \hbox{Re}\,\Sigma^*_2 (\mbox{$\frac{k^2}{2M}$},k)
\,,&\qquad& \gamma_{k,2} = \hbox{Im}\,\Sigma^*_2 (\mbox{$\frac{k^2}{2M}$},k) \,.
\label{qp012}
\end{eqnarray}
At ${\cal O}((\kf/\Lambda)^3)$ we find the equation:
\beq
\label{third_order}
\epsilon_{k,3}+i\gamma_{k,3}=\Sigma^{*}_{3}(\mbox{$\frac{k^2}{2M}$},k)
+\Sigma^{*}_{1}
\frac{\partial\Sigma^{*}_{2}(k_0,k)}{\partial k_0}
\biggr|_{k_0=\frac{k^2}{2M}}~.
\eeq
Then the third order contribution to $\epsilon_p$ and $\gamma_p$ is just 
given by the sum of real and imaginary parts of the diagrams, respectively:
\begin{eqnarray}
\epsilon_{k,3}&=&\hbox{Re}\,\Sigma^*_{3(6)}(\mbox{$\frac{k^2}{2M}$},k)
+\hbox{Re}\,\Sigma^*_{3(7)}(\mbox{$\frac{k^2}{2M}$},k)+
\hbox{Re}\,\Sigma^*_{3(8)}(\mbox{$\frac{k^2}{2M}$},k)
+\hbox{Re}\,\Sigma^*_{3(9)}(\mbox{$\frac{k^2}{2M}$},k)~,
\label{eps3}\\
\gamma_{k,3}&=&\hbox{Im}\,\Sigma^*_{3(6)}(\mbox{$\frac{k^2}{2M}$},k)
+\hbox{Im}\,\Sigma^*_{3(7)}(\mbox{$\frac{k^2}{2M}$},k)~.
\label{gam3}
\end{eqnarray} 
The real and imaginary parts of quasiparticle energy and width
in units of the Fermi energy $E_F=\kf^2/2M$
up to order $(k_F/\Lambda)^3$ are shown in Figs.~\ref{fig_epsk} and 
\ref{fig_gamk}, respectively. 
As an example, we have chosen
the parameters $\kf a_s =k_F a_p=k_F r_s= 0.5$ and $g=4$.
The width $\gamma_k$ shows the behavior predicted by the Lehmann 
representation, that is $\gamma_k>0$ for $k<k_F$ and $\gamma_k<0$ for 
$k>k_F$. 
Our results agree qualitatively with the previous incomplete
calculation of Bund and Wajntal \cite{bund} where the particle-hole
and effective range contributions were omitted.
For those contributions for which a detailed comparison is possible, 
quantitative agreement is obtained as well.

Using Eqs.~(\ref{qp012},\ref{eps3},\ref{gam3}),
we can immediately calculate  the third order contributions
to the above mentioned observables. 
As the effective mass $M^{*}$ is proportional to the inverse of the slope 
of the excitation energy at 
the Fermi surface, we need the complete result for $\epsilon_k$
up to order $(k_F/\Lambda)^3$:
\begin{eqnarray}
\frac{M^{*}}{M}&=&\biggl(1+(g-1)\frac{8}{15\pi^2}(k_F a_s)^2(1-7\log 2)
+\frac{(g+1)}{3\pi}(k_F a_p)^3
+\frac{(\kf a_s)^2 \kf r_s}{6\pi}(g-1) \nonumber\\
&&\qquad +0.11(g-1)(g-3)(k_F a_s)^3-
0.15(g-1)(k_F a_s)^3\biggr)^{-1}
\end{eqnarray}
In order to check our calculation, we calculate the third order
contribution to the chemical potential and the energy per particle.
According to Eq.~(\ref{chemical}), the contribution to the chemical 
potential at ${\cal O}((\kf/\Lambda)^3)$ is given by:
\begin{eqnarray}
\mu_{(3)}&=&\frac{k_F^2}{2M}\biggl[(g-1)\frac{4}{15\pi}(k_F a_s)^2 k_F r_s
+(g+1)\frac{8}{15\pi}(k_F a_p)^3 \nonumber\\
&&\qquad +(g-1)\bigl\{0.20+(g-3)0.15\bigr\}(\kf a_s)^3\biggr]~.
\end{eqnarray}
The third order contribution to the energy per particle is then given by:
\begin{eqnarray}
\biggl(\frac{E}{N}\biggr)_{(3)}&=&\frac{k_F^2}{2M}\biggl[(g-1)
\frac{1}{10\pi}(k_F a_s)^2 k_F r_s+(g+1)\frac{1}{5\pi}(k_F a_p)^3
\nonumber\\
&&\quad+(g-1)\bigl\{0.076+(g-3)0.057\bigr\}(k_F a_s)^3\biggr]~,
\end{eqnarray}
which is in good agreement with previous results 
\cite{efimov,baker,bishop,HAMMER00}.

Finally, we also calculate the wave function renormalization
$Z$ defined in Eq.~(\ref{eq:wfrenorm}).\footnote{Note that $Z$
depends on the representation of fields and is not an observable
in the usual sense. See Ref.~\cite{FuHa02} for a related discussion on the 
representation dependence of the momentum distribution.
We quote $Z$ in the standard representation of fields in which no 
interaction terms proportional to the classical equation of motion 
are included. This representation is equivalent to the Hamiltonian used in
Ref.~\cite{SARTOR80}.} 
It is related to the discontinuity in the momentum distribution at the 
Fermi surface \cite{Migdal57}.
We obtain for $Z(k_F)$:
\begin{eqnarray}
Z(k_F)^{-1}&=&1-\frac{4(g-1)}{\pi^2} \log 2\, (k_F a_s)^2\nonumber\\
&&\qquad+\frac{8 (g-1)}{\pi^3} \{0.184-(g-3)\, 0.224\}(k_F a_s)^3 ~.
\end{eqnarray}
We used the results of Sartor and Mahaux \cite{SARTOR80}
for the occupation numbers to obtain
an analytical expression for the second order term. We verified this
result numerically. At third order, the results were computed numerically.  
\section{Summary and Conclusions}
\label{sec_summary}
In this paper, we have performed the first complete calculation
of the the quasiparticle excitation energies and widths 
in the dilute Fermi gas to order $(k_F/\Lambda)^3$. 
In order to check our calculation, we have derived
and compared the chemical potential and the energy per particle at
$(k_F/\Lambda)^3$ to previous results. 

The calculation has been carried out using an EFT for dilute
Fermi systems \cite{HAMMER00}. Using dimensional regularization with 
minimal subtraction the power counting in this EFT is particularly simple.
All loop momenta are converted to Fermi momenta which leads to a fully 
perturbative expansion. This is in contrast to previous calculations
\cite{efimov,baker,bishop}. There, a more general set of diagrams 
had to be first summed and  then expanded with care to avoid double counting. 
The EFT approach is controlled and allows for systematic improvements
by simply going to the next order in the power counting. Errors can be
estimated reliably using dimensional analysis and error plots.

To order $(\kf/\Lambda)^3$ , no three-body input is required.
At ${\cal O}((\kf/\Lambda)^4)$, however, a contact three-body interaction
enters. Without the three-body term, the logarithmic divergence
appearing at this order can not be renormalized \cite{HAMMER00}. 
A complete calculation 
of the energy density to this order is in progress \cite{Luc02}.

The extension of the present EFT to the nuclear matter problem is
not straightforward. The breakdown scale is set by the lowest momentum
scale in the effective range expansion which is of order 10 MeV for
nuclear matter because of the large S-wave scattering lengths.
Similar problems arise for atomic systems near a Feshbach resonance.
For scattering in the vacuum, the appropriate power counting to deal
with the large scattering lengths is known \cite{vanKolck:1998bw,ksw}. 
It requires to sum
all $C_0$ interactions which for two-particle nonrelativistic scattering
simply involves summing a geometric series. At finite density,
the diagrams involving only $C_0$ vertices are not easily summed
because both particles and holes are present. To make the problem
tractable, an additional ordering scheme such as the hole-line
expansion in the standard nuclear physics approach is needed. 
Whether the hole-line expansion can be justified in EFT using
power counting arguments remains to be seen. An interesting 
possibility is to exploit the geometry of intersecting 
Fermi spheres inherent in the finite density loop integrals which
leads to an expansion in $1/D$ where $D$ is the space time dimension
\cite{STEELE00}. Furthermore, an expansion in the inverse spin-isospin
degeneracy $g=4$ might prove useful \cite{FURNSTAHL02}.
The nuclear matter problem provides additional challenges.
Since $\kf\approx 270$ MeV, the pion exchange is long range and cannot 
be treated within a purely short range EFT. Both the large scattering length
problem and the inclusion of the long range pion exchange are important
questions to be addressed by future work.

Recent progress in the field of cold Fermi gases \cite{jin99,hulet01}
also opens the possibility of applications of EFT methods in the 
physics of cold fermionic atoms. In the case of cold bosons,
EFT methods have already been applied successfully to a number of
processes \cite{Bedaque:2000ft,Braaten:2001hf,Braaten:2001ay,Bedaque:2002xy}.


\acknowledgments

We thank R.J.\ Furnstahl and A.~Schwenk for useful discussions.
HWH thanks the Benasque Center for Science for its hospitality
and partial support during completion of this work.
This work was supported in part by the U.S. National Science
Foundation under Grant No.\ PHY--0098645.
\pagebreak

\begin{appendix}

\section{Proper Self-Energy Integrals}
\label{app_integration}
We illustrate the calculation of the self-energy diagrams
in Section \ref{sec_selfenergy} by applying the 
Feynman rules to diagram (7) from Fig.~\ref{fig_sigma3}. This 
diagram has a symmetry factor of $1/4$. Thus, one obtains 
the following expression:
\beq
\Sigma^*_{3(7)}(\tilde p)=(-i C_0)^3\frac{i}{4}\frac{4(1-g)}{(2\pi)^{12}}
\int\hbox{d}^4 k \int\hbox{d}^4 q\int\hbox{d}^4 l \; i^5 G_0(\tilde k)
G_0(\tilde p-\tilde q)G_0(\tilde k+\tilde q)G_0(\tilde p-\tilde l)~.
\eeq
After performing the contour integrations we have:
\begin{eqnarray}
\Sigma^*_{3(7)}(\tilde p)&=&C_0^3 \frac{g-1}{(2\pi)^9}\int\hbox{d}^3 k 
\int\hbox{d}^3 q\int\hbox{d}^3 l\nonumber\\
&&\Biggl[\frac{2\theta(k-k_F)\theta(|\mathbf{k}+\mathbf{l}|-k_F)
\theta(|\mathbf{p}-\mathbf{l}|-k_F)\theta(k_F-|\mathbf{k}+\mathbf{q}|)
\theta(k_F-|\mathbf{p}-\mathbf{q}|)}{(p_o+\omega_k-\omega_{\mathbf{k}+
\mathbf{q}}-\omega_{\mathbf{p}-\mathbf{q}}-i\epsilon)(\omega_{\mathbf{k}+
\mathbf{q}}-\omega_{\mathbf{k}+\mathbf{l}}-\omega_{\mathbf{p}-\mathbf{l}}+
\omega_{\mathbf{p}-\mathbf{q}}+i\epsilon)}\nonumber\\
&&- \frac{\theta(k-k_F)\theta(k_F-|\mathbf{k}+\mathbf{l}|)
\theta(k_F-|\mathbf{p}-\mathbf{l}|)\theta(k_F-|\mathbf{k}+\mathbf{q}|)
\theta(k_F-|\mathbf{p}-\mathbf{q}|)}{(p_o+\omega_k-\omega_{\mathbf{k}+
\mathbf{l}}-\omega_{\mathbf{p}-\mathbf{l}}-i\epsilon)(p_o+\omega_k-
\omega_{\mathbf{k}+\mathbf{q}}-\omega_{\mathbf{p}-\mathbf{q}}-i\epsilon)}
\label{eq:a2}\\
&&-\frac{2\theta(k_F-k)\theta(|\mathbf{k}+\mathbf{l}|-k_F)
\theta(|\mathbf{p}-\mathbf{l}|-k_F)\theta(k_F-|\mathbf{k}+\mathbf{q}|)
\theta(k_F-|\mathbf{p}-\mathbf{q}|)}{(p_o+\omega_k-\omega_{\mathbf{k}+
\mathbf{l}}-\omega_{\mathbf{p}-\mathbf{l}}+i\epsilon)(\omega_{\mathbf{k}+
\mathbf{l}}-\omega_{\mathbf{k}+\mathbf{q}}+\omega_{\mathbf{p}-\mathbf{l}}-
\omega_{\mathbf{p}-\mathbf{q}}-i\epsilon)}\nonumber\\
&&+\frac{\theta(k_F-k)\theta(|\mathbf{k}+\mathbf{l}|-k_F)\theta(|\mathbf{p}-
\mathbf{l}|-k_F)\theta(|\mathbf{k}+\mathbf{q}|-k_F)\theta(|\mathbf{p}-
\mathbf{q}|-k_F)}{(p_o+\omega_k-\omega_{\mathbf{k}+\mathbf{l}}-\omega_{
\mathbf{p}-\mathbf{l}}+i\epsilon)(p_o+\omega_k-\omega_{\mathbf{k}+
\mathbf{q}}-\omega_{\mathbf{p}-\mathbf{q}}+i\epsilon)}\Biggr]\nonumber~.
\end{eqnarray}
By setting this expression on-shell (that is $p_0=p^2/(2M)$)
and an appropriate substitution to dimensionless variables using
\beq
\mathbf{k}=k_F \mathbf u,\quad\mathbf{q}=k_F\frac{1}{2}(
\mathbf{v} -\mathbf{u} -2\mathbf{t}),\quad\mathbf{l}=k_F\frac{1}{2}(
\mathbf{v} -\mathbf{u} -2\mathbf{r}), \quad\mathbf{p}=k_F \mathbf{v}~,
\eeq 
Eq.~(\ref{eq:a2}) can be simplified significantly:
\begin{eqnarray}
\Sigma^*_{3(7)}(\tilde p)
&=&\frac{C_0^3 M^2 k_F^5(g-1)}{(2\pi)^9}\int\hbox{d}^3u \int\hbox{d}^3t
\int\hbox{d}^3r\nonumber\\
&&\Biggl[\frac{2\theta(u-1)\theta(|\mathbf{s}-\mathbf{r}|-1)\theta(|
\mathbf{s}-\mathbf{r}|-1)\theta(1-|\mathbf{s}-\mathbf{t}|)\theta(1-|
\mathbf{s}+\mathbf{t}|)}{(2q^2-2t^2-i\epsilon)(2t^2-2r^2-i\epsilon)}\nonumber\\
&&-\frac{\theta(u-1)\theta(1-|\mathbf{s}-\mathbf{r}|)\theta(1-|\mathbf{s}-
\mathbf{r}|)\theta(1-|\mathbf{s}-\mathbf{t}|)\theta(1-|\mathbf{s}+
\mathbf{t}|)}{(2q^2-2r^2-i\epsilon)(2q^2-2t^2-i\epsilon)}
\label{eq:a4}\\
&&-\frac{2\theta(1-u)\theta(|\mathbf{s}-\mathbf{r}|-1)\theta(|\mathbf{s}-
\mathbf{r}|-1)\theta(1-|\mathbf{s}-\mathbf{t}|)\theta(1-|\mathbf{s}+
\mathbf{t}|)}{(2q^2-2r^2-i\epsilon)(2r^2-2t^2-i\epsilon)}\nonumber\\
&&+\frac{\theta(1-u)\theta(|\mathbf{s}-\mathbf{r}|-1)\theta(|\mathbf{s}-
\mathbf{r}|-1)\theta(|\mathbf{s}-\mathbf{t}|-1)\theta(|\mathbf{s}+\mathbf{t}|
-1)}{(2q^2-2r^2+i\epsilon)(2q^2-2t^2+i\epsilon)}\Biggr]~.\nonumber
\end{eqnarray}
We consider only the first term in the square brackets:
\begin{eqnarray}
\mathcal{I}_1=\int\hbox{d}^3u \int\hbox{d}^3t\int\hbox{d}^3r \frac{2\theta(u-1)
\theta(|\mathbf{s}-\mathbf{r}|-1)\theta(|\mathbf{s}-\mathbf{r}|-1)
\theta(1-|\mathbf{s}-\mathbf{t}|)\theta(1-|\mathbf{s}+\mathbf{t}|)}
{(2q^2-2t^2-i\epsilon)(2t^2-2r^2-i\epsilon)}~,
\end{eqnarray}
where the constant overall prefactor in Eq.~(\ref{eq:a4}) has been
dropped. By analyzing the step function, we can rewrite the integral as
\begin{eqnarray}
\mathcal{I}_1=8\pi^3 \int_{-1}^{1}\hbox{d}x\int_{1}^{\infty}\hbox{d}u
\int_{0}^{1}\hbox{d}z\int_{0}^{z_{-}(s,z)}\hbox{d}t
\int_{0}^{1}\hbox{d}y\int_{z_{+}(s,y)}^{\infty}\hbox{d}r
\frac{u^2 t^2 r^2 \theta(1-s)}{(q^2-t^2-i\epsilon)(t^2-r^2+i\epsilon)}~,
\end{eqnarray}
with
\begin{equation}
z_{\pm}(s,z)=\pm s z +\sqrt{1-s^2(1-z^2)}~. 
\end{equation}
The integral over $r$ contains a linearly divergent term which can be 
separated by writing
\begin{equation}
\int_{z_{+}(s,y)}^{\infty}\hbox{d}r\frac{r^2 }{t^2-r^2+i\epsilon}=
\int_{0}^{\infty}\hbox{d}r\frac{r^2 }{t^2-r^2+i\epsilon}+
\int_{0}^{z_{+}(s,y)}\hbox{d}r\frac{r^2 }{t^2-r^2+i\epsilon}~.
\label{eq:a8}
\end{equation}
The ultraviolet divergence is contained in the first term 
on the right-hand side of Eq.~(\ref{eq:a8}). 
In dimensional regularization with minimal subtraction
the linear divergence is subtracted automatically and the finite part
of this term is purely imaginary. Thus, we can write $\mathcal{I}_1$ as:
\begin{eqnarray}
\mathcal{I}_1&=&8\pi^3 \int_{-1}^{1}\hbox{d}x\int_{1}^{\infty}\hbox{d}u
\int_{0}^{1}\hbox{d}z\int_{0}^{z_{-}(s,z)}\hbox{d}t\quad
\frac{u^2 t^2 \theta(1-s)}{(q^2-t^2-i\epsilon)}\biggl(-\frac{i\pi t}{2}
\biggr)\\
&&-8\pi^3 \int_{-1}^{1}\hbox{d}x\int_{1}^{\infty}\hbox{d}u
\int_{0}^{1}\hbox{d}z\int_{0}^{z_{-}(s,z)}\hbox{d}t
\int_{0}^{1}\hbox{d}y\int_{0}^{z_{+}(s,y)}\hbox{d}r
\frac{u^2 t^2 r^2 \theta(1-s)}{(q^2-t^2-i\epsilon)(t^2-r^2+i\epsilon)}~.
\nonumber
\end{eqnarray}
Separating the real and imaginary part of this integral and some further 
straightforward manipulations lead to expressions that can be integrated 
numerically. The remaining terms in the square brackets 
in Eq.~(\ref{eq:a4}) are treated analogously.

\end{appendix}

\end{document}